\documentclass[journal]{IEEEtran}

\usepackage{acronym}
\usepackage{amsfonts}
\usepackage[dvips]{graphicx}
\usepackage{times}
\usepackage{cite}
\usepackage{amsmath}
\usepackage{array}
\usepackage{amssymb}
\usepackage{stfloats}
\usepackage{diagbox}
\usepackage{graphicx}
\usepackage{footnote}
\usepackage{amsthm}
\usepackage{booktabs}
\usepackage{array}
\usepackage[ruled,vlined]{algorithm2e}
\usepackage{subeqnarray}
\usepackage{cases}
\usepackage{threeparttable}
\usepackage{color}
\usepackage{epstopdf}
\usepackage{multirow}
\usepackage{tabularx}
\usepackage{enumerate}
\usepackage{multicol}
\usepackage{hyperref}
\usepackage{subfig}
\usepackage{caption}

\def\bb0{{\mathbb{0}}}

\def\ba{{\mathbf{a}}}
\def\bb{{\mathbf{b}}}
\def\bc{{\mathbf{c}}}
\def\bd{{\mathbf{d}}}

\def\bg{{\mathbf{g}}}

\def\bm{{\mathbf{m}}}
\def\bn{{\mathbf{n}}}

\def\br{{\mathbf{r}}}

\def\bu{{\mathbf{u}}}
\def\bv{{\mathbf{v}}}

\def\bx{{\mathbf{x}}}
\def\by{{\mathbf{y}}}
\def\bz{{\mathbf{z}}}
\def\b0{{\mathbf{0}}}

\def\bA{{\mathbf{A}}}

\def\bE{{\mathbf{E}}}

\def\bH{{\mathbf{H}}}
\def\bI{{\mathbf{I}}}

\def\bW{{\mathbf{W}}}

\def\bbC{{\mathbb{C}}}

\def\bbR{{\mathbb{R}}}

\def\cF{\mathcal{F}}

\def\cN{\mathcal{N}}

\def\cS{\mathcal{S}}

\def\sfD{\mathsf{D}}

\def\sfP{\mathsf{P}}

\def\sfR{\mathsf{R}}

\def\sfU{\mathsf{U}}

\def\sfee{{\mathsf{e}}}

\def\sfp{{\mathsf{p}}}
\def\sfq{{\mathsf{q}}}

\def\sfx{{\mathsf{x}}}

\def\sf0{{\mathsf{0}}}

\def\rmd{{\mathrm{d}}}

\def\rm0{{\mathrm{0}}}

\def\j{\mathrm{j}}

\acrodef{CSI}[CSI]{channel state information}
\acrodef{CSIT}[CSIT]{channel state information at the transmitter}
\acrodef{CSIR}[CSIR]{channel state information at the receiver}
\acrodef{MIMO}[MIMO]{multiple-input multiple-output}
\acrodef{SISO}[SISO]{single-input single-output}
\acrodef{MISO}[MISO]{multiple-input single-output}
\acrodef{SIMO}[SIMO]{single-input multiple-output}
\acrodef{ADCs}[ADCs]{analog-to-digital convertors}
\acrodef{SNR}[SNR]{signal-to-noise ratio}
\acrodef{AWGN}[AWGN]{additive white Gaussian noise}
\acrodef{MRT}[MRT]{maximal ratio transmission}
\acrodef{DFT}[DFT]{Discrete Fourier Transform}
\acrodef{ULA}[ULA]{uniform linear array}
\acrodef{UPA}[UPA]{uniform planar array}
\acrodef{LS}[LS]{least squares}
\acrodef{ALMMSE}[ALMMSE]{approximate linear minimum mean squared error}
\acrodef{QIHT}[QIHT]{quantized iterative hard thresholding}
\acrodef{QIST}[QIST]{quantized iterative soft thresholding}
\acrodef{SVD}[SVD]{singular value decomposition}

\ifCLASSINFOpdf

\else
\fi
\begin{document}
\title{Message Passing Meets Graph Neural Networks: A New Paradigm for Massive MIMO Systems}
\author{Hengtao He, \IEEEmembership{Member,~IEEE,}
Xianghao Yu, \IEEEmembership{Member,~IEEE,}
Jun Zhang,~\IEEEmembership{Fellow,~IEEE,}
Shenghui Song,~\IEEEmembership{Senior Member,~IEEE,}
and Khaled B. Letaief,~\IEEEmembership{Fellow,~IEEE}

\thanks{Manuscript received February 14, 2023; revised July 14, 2023; accepted September 23, 2023. This paper was presented in part at the IEEE Wireless Commun. Netw. Conf. (WCNC), Glasgow, Scotland, UK, 2023 \cite{AMP_GNN}. This work was supported in part by the Hong Kong Research Grants Council under the Areas of Excellence Scheme Grant AoE/E-601/22-R.  The work of Shenghui Song was supported by a grant from the NSFC/RGC Joint Research Scheme sponsored by the Research Grants Council of the Hong Kong Special Administrative Region, China and National Natural Science Foundation of China (Project No. N\_HKUST656/22). This work of Jun Zhang was supported in part by the Hong Kong Research Grants Council under Grant 16209622. The work of Xianghao Yu was supported by the Hong Kong Research Grants Council under Grant No. 16212922 and City University of Hong Kong under Project No. 9610629. The associate editor coordinating the review of this paper and approving it for publication was Prof. Pengfei Hu. \emph{(Corresponding author: Xianghao Yu.)}}

\thanks{H. He, J. Zhang, S. Song, and K. B. Letaief are with the Department of Electronic and Computer Engineering,
the Hong Kong University of Science and Technology, Hong Kong, E-mail: \{eehthe, eejzhang, eeshsong, eekhaled\}@ust.hk.}
\thanks{Xianghao Yu is with the Department of Electrical Engineering, City University of Hong Kong (CityU), Hong Kong (e-mail: alex.yu@cityu.edu.hk).}

}

\maketitle

\begin{abstract}

As one of the core technologies for 5G systems, massive multiple-input multiple-output (MIMO) introduces dramatic capacity improvements along with very high beamforming and spatial multiplexing gains. When developing efficient physical layer algorithms for massive MIMO systems,  message passing is one promising candidate owing to its  superior performance. However, as their computational complexity increases dramatically with the problem size, the state-of-the-art message passing algorithms cannot be directly applied to future 6G systems, where an exceedingly large number of antennas are expected to be deployed. To address this issue, we propose a  model-driven deep learning (DL) framework, namely the AMP-GNN for massive MIMO transceiver design, by considering the \emph{low complexity} of the AMP algorithm and \emph{adaptability} of GNNs. Specifically, the structure of the AMP-GNN network is customized by unfolding the approximate message passing (AMP) algorithm and introducing a graph neural network (GNN) module into it.  The permutation equivariance property of AMP-GNN is proved, which enables the AMP-GNN to learn more efficiently and to adapt to different numbers of users. We also reveal the underlying reason why GNNs improve the AMP algorithm from the perspective of  expectation propagation, which motivates us to amalgamate various GNNs with different message passing algorithms. In the simulation, we take the massive MIMO detection to exemplify  that the proposed AMP-GNN significantly improves the performance of the AMP detector, achieves comparable performance as the state-of-the-art DL-based MIMO detectors, and presents strong robustness to various mismatches.

\end{abstract}

\begin{IEEEkeywords}
6G, Bayesian inference, deep learning, graph neural networks, massive MIMO, model-driven.
\end{IEEEkeywords}

%
\IEEEpeerreviewmaketitle

\section{Introduction}
Massive multiple-input multiple-output (MIMO) has been proven as one of the crucial enabling technologies for fifth-generation (5G) systems, where the transmitter/receiver is equipped with tens or even hundreds of antennas to improve the system throughput and spectral efficiency dramatically \cite{massiveMIMO}. It  overcomes many challenges such as the massive data traffic and users, large free-space path-loss in millimeter-wave (mmWave) systems, etc. It has been first commercialized with 64-antenna massive MIMO base stations widely deployed by Ericsson, Nokia AirScale, and Huawei \cite{product}.

As the adoption of the 5G wireless networks continues to accelerate around the world \cite{5G}, we are witnessing exciting global research and development activities to formulate the next-generation mobile communication network. This trend is reinforced by the recent emergence of several innovative applications, including the Internet of Everything, Tactile Internet, and seamless virtual and augmented reality \cite{6G_khaled}. Future wireless networks (6G) are expected to provide ubiquitous coverage, enhanced spectral efficiency (SE), connected intelligence, etc \cite{Khlaed_JSAC}. Facing the new demand, massive MIMO will continuously evolve to ultra-massive MIMO where hundreds or even thousands of antennas are deployed in centralized or distributed manners, and keep playing important roles in future 6G wireless networks \cite{massiveMIMOreality}. Given the increasing size of antenna arrays, the computational complexity is one of the critical issues when developing efficient physical layer  algorithms, such as channel estimation and MIMO detection. In essence, these transceiver design problems can be categorized as  \emph{high-dimensional statistical inference} problems and several approaches have been developed accordingly \cite{MIMOWireless}.
\subsection{Related Works}
The high-dimensional statistical inference problem can be solved by exploiting probabilistic graphical models (PGMs) \cite{PRML}. Based on the PGMs, many approximate inference algorithms have been developed, including belief propagation (BP)\cite{factor_graph}, approximated message passing (AMP)\cite{dynamicAMP}, and expectation propagation (EP)\cite{EP}. These algorithms consider different iterative approaches and have been widely applied to physical layer processing in wireless communications \cite{AMPJSTSP,EPdetector}. For instance, iterative detectors based on AMP and EP have been proposed \cite{AMPJSTSP,EPdetector}. The AMP-based detector \cite{AMPJSTSP} is of low complexity and easy to implement in practice because only the matrix-vector multiplication is involved. In contrast, the EP-based detector \cite{EPdetector} achieves Bayes-optimal performance when the channel matrix is unitarily invariant. Nevertheless, it has an extremely higher complexity than the AMP-based detector owing to the required matrix inversion. On the other hand, the AMP algorithm has been applied for massive MIMO channel estimation by exploiting the sparsity of the channel observed in the beam domain \cite{gaussian_mixture}. 
A similar idea has been considered by adopting the EP algorithm with better performance but higher complexity \cite{EPCE}. Furthermore, AMP and EP algorithms have  been applied to coded linear systems and show asymptotically optimal performance \cite{AMPcoded, OAMPcoded}. These applications verify the inherent low-complexity of the AMP algorithm, which is more promising for
future wireless communications equipped with large-scale antennas.

Thanks to the strong ability of extracting representative features from data, deep learning (DL) has been recently utilized in the physical layer design of wireless communications \cite{DL2017shea,DL2018Qin, Modeldriven18DL,DL2021Hoydis}, such as mmWave channel estimation \cite{DL2018HE, DL2022HE}, channel state information (CSI) feedback \cite{DL2018Wen}, and data detection \cite{DL2OFDM,MIMO_Detection20DL,DeepRX,REMIMO20DL}. Given the interpretability of model-based DL, researchers started to make efforts on improving the message passing-based algorithms in wireless communications by utilizing the model-driven DL technology \cite{Modeldriven18DL}. Among these works, incorporating the learnable modules into the message passing algorithms is a promising way. For example, convolutional neural network (CNN)-based denoiser has been introduced into the AMP and generalized expectation consistent signal recovery (GEC-SR) algorithms for narrow and wideband beamspace channel estimation, respectively\cite{DL2018HE, DL2022HE}.  They both improved the message passing-based algorithms and  achieved excellent performance even with a small number of RF chains. Furthermore, the orthogonal AMP (OAMP)-Net and OAMP-Net2 detectors \cite{MIMO_Detection20DL} were developed by unfolding the OAMP detector \cite{OAMP} and introducing several learnable parameters. Such detectors were shown to achieve a significant performance improvement compared with the OAMP detector due to the learnable parameters. On the other hand, it has been shown that DL methods can be used to improve a standard BP decoder by assigning learnable weights to the edges in the Tanner graph \cite{DLBP}. Although these works have demonstrated  performance improvement by introducing DL, they are trained for fixed MIMO configurations and suffer from poor generalization to handle varying numbers of users or antennas with a single model \cite{DLBP}. This is because these networks have fixed dimensions of the input and output, and lack the inherent structure to adapt to the dynamic dimension of the problem.

To solve this problem, structured neural networks have been introduced to the design of wireless communications. As one of  these networks, graph neural networks (GNNs) have attracted much attention in the machine learning field \cite{Yoon18,Gilmer17,GNN_Survey} because of their flexible and adaptive structure. They have  been recently adopted in \cite{Unfolding_WMMSE,NEBP,YifeiGNN1,YifeiGNN2,Scotti20GNN} to exploit the domain knowledge and have been shown to generalize well to different system settings. By incorporating the graph topology of the wireless network into the neural network design, they can improve scalability and generalization. 
Moreover, GNNs have been applied to learn a message passing solution for  statistical inference problems \cite{Scotti20GNN,Kosasih20DL}.
In particular, a GNN-based MIMO detector was developed by utilizing a pair-wise Markov random field (MRF) model \cite{Scotti20GNN}. However, the performance is far worse than existing message passing detectors. More recently, the GEPNet was developed by incorporating the GNN into the EP detector\cite{Kosasih20DL,Zhou22DL}. Although it can achieve the state-of-the-art performance, the computational complexity is prohibitively high because of the matrix inversion in each layer, which is not affordable in future wireless networks with a larger number of antennas. By far, an efficient transceiver design framework, which strikes a better balance between performance and complexity and adapts to  dynamic system configurations, is  not available.
\subsection{Contributions}
To fill this gap, we develop a \emph{low-complexity} model-driven DL-based framework, namely AMP-GNN, which benefits from both  the low-complexity of  AMP  and the adaptability of GNNs. By leveraging the deep unfolding technique \cite{Deep19Eldar,deepunfolding}, we construct the network structure by unfolding the AMP algorithm and incorporating the GNNs module.  The main contributions of this work are summarized as follows.
\begin{itemize}
  \item We first introduce several statistical inference problems and  DL-enhanced message passing algorithms in physical layer communications.
     By incorporating the message passing neural network (MPNN)\footnote{MPNN is one of the representative GNNs and can unify various GNNs proposed in \cite{Gilmer17}.} into the AMP algorithm, we propose a model-driven DL framework, namely AMP-GNN,  for massive MIMO transceiver design. In particular, the MPNN module receives the equivalent additive white Gaussian noise (AWGN) observations from  AMP as the input and outputs a refined version back to AMP at each layer. Thus, it inherits the low complexity  advantages of the AMP  algorithm and adaptability from GNNs, which are  desirable features for massive MIMO systems.

  \item We prove that AMP-GNN has the property of permutation equivalence, which is favorable for learning more efficiently, avoiding over-fitting, and
        developing strong generalizability. Furthermore, we reveal the reason why MPNN can improve the AMP algorithm from the perspective of the EP. This will then provide us with insightful guidelines to combine various GNNs with different message passing algorithms.
  \item We take the massive MIMO detection as an application and demonstrate that the proposed AMP-GNN-based detector significantly outperforms the existing AMP detector. Furthermore, the AMP-GNN-based detector entails a much lower computational complexity yet comparable performance compared to the state-of-the-art GEPNet detector.  Simulation results also show that the AMP-GNN-based detector is robust to channel estimation errors and generalize to different numbers of users with a single model.
\end{itemize}

\emph{Notations}---For any matrix $\mathbf{A}$, $\mathbf{A}^{T}$, $\mathbf{A}^{*}$, and ${ \mathrm{tr}}(\mathbf{A})$  denote the transpose, conjugate, and  trace of $\mathbf{A}$, respectively. In addition, $\mathbf{I}$ is the identity matrix, $\mathbf{0}$ is the zero matrix, and $\mathbf{1}_n$ is the $n$-dimensional all-ones vector. A proper complex Gaussian distribution with mean $\boldsymbol{\mu}$ and covariance $\boldsymbol{\Omega}$ can be described by the probability density function (pdf):
\begin{equation*}
  \mathcal{N}_{\bbC}(\bz;\boldsymbol{\mu},\boldsymbol{\Omega})=\frac{1}{\mathrm{det}(\pi \boldsymbol{\Omega})}
  e^{-(\bz-\boldsymbol{\mu})^{H}\boldsymbol{\Omega}^{-1}(\bz-\boldsymbol{\mu})}.
\end{equation*}
The remaining part of this paper is organized as follows. Section \ref{Problem}
identifies classical statistical inference problems and reviews existing DL-based message passing  algorithms. Next, the AMP-GNN is proposed in Section \ref{Sec:AMP_GNN} and several key properties are also analyzed.  Section \ref{Applications} elaborates the AMP-GNN for massive MIMO detection. Numerical results are then presented in Section \ref{Simulation}. Finally, Section \ref{con} concludes the paper.
\section{Problem Formulation and Algorithms Review}\label{Problem}
In this section, we first introduce the statistical inference problems in wireless communications in detail. Then, these problems are formulated under the framework of  Bayesian inference and the factor graph is elaborated to characterize the inference problem for deriving efficient algorithms. Finally, several existing message passing algorithms are reviewed.
\subsection{Statistical Inference Problems in Wireless Communications} \label{Sec:inference}
In wireless communication systems, statistical inference is widely utilized in many areas, ranging from transceivers design to network optimization. We consider a standard statistical inference problem that aims to recover the unknown signal $\bx \in \bbC^{N} $ from a set of measurements $\by \in \bbC^{M}$ with the system model
\begin{equation}\label{eqfirst}
  \by=\bA \bx+\bn,
\end{equation}
where $\bA \in \bbC^{M\times N}$  is the measurement matrix and $\bn$ is the additive white Gaussian noise (AWGN). In particular, we list several typical statistical problems in the wireless physical layer in Table\,\ref{statistical_inference}, including channel estimation, MIMO detection, finite-alphabet precoding, and channel decoding. When developing an efficient algorithm to recover signal $\bx$, it is assumed that prior knowledge of the underlying signal is known \cite{BayesianCS}. Such prior knowledge is typically based on the assumption that $\bx$ lies in a restricted set $\cS$, or follows some distribution with known/unknown parameters. For example, the transmitted symbols $\bx$ are the discrete signals chosen from the $Q$-QAM set $\cS$ and the channel is assumed to be Gaussian-mixture distributed with unknown statistical parameters.  To solve the statistical inference problem in \eqref{eqfirst}, the Bayesian estimator is considered  an efficient approach, which will be introduced in the next subsection.
\begin{table*}[t]
\renewcommand{\arraystretch}{1.4}
\caption{Examples of statistical inference problems in physical layer communications.}
\label{statistical_inference}
\centering 
\begin{tabular}{|c|c|c|c|}
\hline
\hline
\textbf{Problem} & \textbf{Estimated signals $\bx$}  & \textbf{Measurement matrix $\bA$} & \textbf{Measurements $\by$} \\
\hline
Channel estimation & Channel & Pilots & Received pilot signals \\
\hline
MIMO detection & Transmitted signals & MIMO channel & Received data signals \\
\hline
Finite-alphabet precoding & Precoding signals & Channel & Transmitted signals \\
\hline
Channel decoding & Codewords & Identity matrix &  Detected symbols \\
\hline
\end{tabular}
\end{table*}
\subsection{Bayesian Inference and Factor Graph}\label{sec:Bayesian}
When solving the statistical inference problem, a powerful  approach is to use  probabilistic  inference under the framework of the Bayesian methodology. According to Bayes' theorem, the posterior probability $\sfP(\bx|\by,\bA)$ can be factorized as
\begin{equation}\label{eq_posterior}
    \sfP(\bx|\by,\bA)=\frac{\sfP(\by|\bx,\bA)\sfP(\bx)}{\sfP(\by|\bA)}
 = \frac{\sfP(\by|\bx,\bA)\sfP(\bx)}{\int \sfP(\by|\bx,\bA)\sfP(\bx)\rmd\bx}.
\end{equation}
Given the posterior probability $\sfP(\bx|\by,\bA)$, the Bayesian MMSE estimate is obtained by
\begin{equation}\label{MMSE_estimate}
\hat{\bx}=\int \bx \sfP(\bx |\by,\bA) \rmd \bx.
\end{equation}
%
However, the Bayesian MMSE estimator is often intractable because the marginal posterior probability for each element in (\ref{MMSE_estimate}) involves a high-dimensional integral, which motivated researchers to develop the approximate inference to obtain the marginal posterior probability effectively. 
Factor graph, visualizing the dependency on a set of variables with a bipartite graph, is a useful method to develop approximate inference algorithms. It can simplify a joint probability distribution $\sfP(\bx|\by,\bA)$ over many variables $\bx$ by factorizing the distribution according to conditional independence relationships.

\begin{figure}[t]
  \centering
  \includegraphics[width=8cm]{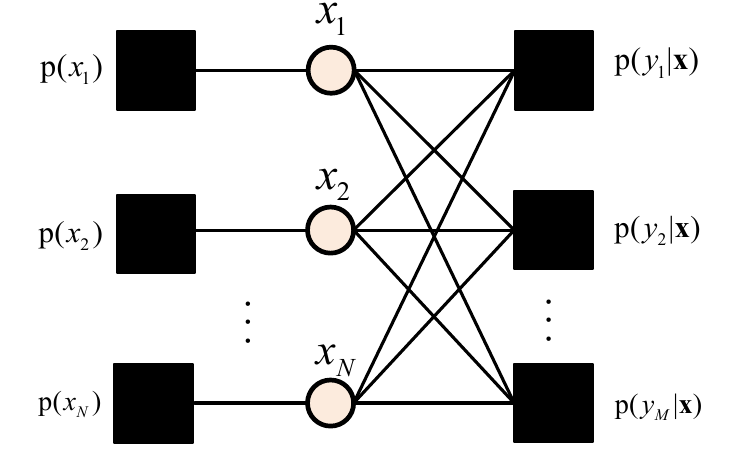}
  \caption{\small{.~~Factor graph for inverse problems.}}\label{fig:factor}
\end{figure}

As illustrated in Fig.\,\ref{fig:factor}, the factor graph consists of two kinds of nodes, where the hollow circles represent the variable nodes and the solid squares represent the factor nodes. Based on the factor graph, efficient message passing algorithms for solving inference problems can be obtained by performing different rules. One of the well-known iterative inference algorithms is the BP, which is denoted by the following equations,
\begin{subequations}
\begin{align}
\mu_{n\rightarrow m}^{(t+1)}(x_n)&\propto  \sfP(x_n)\prod_{b}^M \mu_{b \rightarrow n}^{(t)}(x_n),\\
\mu_{m\rightarrow n}^{(t)}(x_n)&\propto \int \sfP(y_m|\bx)\prod_{j\ne n}^N\mu_{j\rightarrow a}^{(t)}(x_j)\text{d}\bx_{\backslash n},
\label{Chapter_BP}
\end{align}
\end{subequations}
\noindent where the superscript $t$ denotes the index of iteration, $\bx_{\backslash n}$ denotes the vector composed by removing the element $x_{n}$ from $\bx$, $\mu_{n\rightarrow m}^{(t+1)}(x_n)$ is the message from variable node $x_n$ to factor node $\sfP(y_m|\bx)$ at the $(t+1)$-iteration, and $\mu_{n\leftarrow m}^{(t)}(x_n)$ is the message in the opposite direction. Note that the marginal posterior $\sfP(x_n|\by)$ at the $t$-iteration can be approximated by
\begin{align}
\hat{\sfP}^{(t+1)}(x_n|\by)=\frac{\sfP(x_n)\prod_{m=1}^M \mu_{n\leftarrow m}^{(t)}(x_n)}{\int \sfP(x_n) \prod_{m=1}^M \mu_{n\leftarrow m}^{(t)}(x_n) \text{d}x_n}.
\label{Equ:posterior}
\end{align}
Thus, the mean of the approximated posterior $\hat{\sfP}^{(t+1)}(x_n|\by)$ can serve as the result of the Bayesian MMSE estimator. 
Next, we will introduce several classical  message passing algorithms and summarize them under a general framework.
\subsection{Existing Message Passing Algorithms}\label{sec:AMP}
Each iteration of existing message passing algorithms can be represented by the following iterative framework
\begin{subequations}\label{eqtwo}
\begin{align}
\mathrm{LM}: \quad \br^{(t)} &= \gamma_t (\hat{\bx}^{(t)},\hat{\bv}^{(t)}),\\
\mathrm{NLM}:  \quad   \hat{\bx}^{(t+1)} &= \eta_t ( \br^{(t)},\boldsymbol{\Sigma}^{(t)}),
\end{align}
\end{subequations}
where $\mathrm{LM}$ and  $\mathrm{NLM}$ represent the \emph{linear module} $\gamma_{t}$  and the \emph{nonlinear module} $\eta_t$, respectively. The linear module $\gamma_{t}$ takes the current estimate $\hat{\bx}^{(t)}$  and associate variance estimate $\hat{\bv}^{(t)}$  as inputs, and outputs  an intermediate signal $\br^{(t)}$ as well as the associate variance estimate $\boldsymbol{\Sigma}^{(t)}$. In particular, the linear module is a linear transformation for the input $\hat{\bv}^{(t)}$ and $\br^{(t)}$. The function of the linear module is to decouple the linear mixing model (\ref{eqfirst}) into a series of independent equivalent AWGN channels given by
\begin{equation}\label{eqAWGN}
    r_{n}^{(t)} = x_{n} + w_{n}^{(t)},
\end{equation}
where  $w_{n}^{(t)}\sim \cN_{\mathbb{C}}(w_{n}^{(t)};0,\Sigma_n^{(t)})$. On the other hand, the nonlinear module $\eta_{t}$  takes $\br^{(t)}$ and $\boldsymbol{\Sigma}^{(t)}$ as inputs, and utilizes denoising operation\footnote{Denoising operation means the recovery of the signal $x_{n}$ by removing equivalent noise $w_{n}^{(t)}$ from \eqref{eqAWGN}.} for $\br^{(t)}$ to obtain a new estimate $\hat{\bx}^{(t+1)}$.

In essence,  a more powerful linear module can achieve better performance as it can significantly reduce the correlation between different equivalent AWGN channels in (\ref{eqAWGN}).  For example, because of  the LMMSE estimator used in the linear modules in the OAMP and EP algorithms, they can outperform the AMP algorithm. 
On the other hand, DL-based solutions have been recently shown that it can improve traditional message passing algorithms, such as AMP, OAMP, and EP. As illustrated in Table\,\ref{messagepassing}, they have been applied to wireless communications with different strategies to tackle different physical layer design problems. Specifically, the OAMP-Net was developed by unfolding the OAMP algorithm and introducing several learnable parameters \cite{MIMO_Detection20DL}. The idea is using DL to enhance the performance of the linear module with several scalar learnable variables. Furthermore, the GEPNet, proposed in \cite{Kosasih20DL}, unfolds the EP algorithm and uses the GNNs to further improve the performance of the LMMSE linear module. 
However, it has a prohibitively high computational complexity  due to the matrix inversion in LMMSE module. As the antenna size is expected to be extremely large in future ultra-massive MIMO systems, it is of paramount importance to adopt a low-complexity linear module, e.g., match filtering (MF), in the message passing algorithm and seek a sophisticated way to further improve the performance of the detector by capitalizing on GNNs.
\begin{table*}
\renewcommand{\arraystretch}{1.4}
\caption{Components of different message passing Algorithms}
\label{messagepassing}
\centering 
\begin{tabular}{|c|c|c|c|}
\hline
\textbf{Algorithms} & \textbf{Linear module} & \textbf{Nonlinear module} & \textbf{Learnable module}  \\
\hline
AMP \cite{AMPJSTSP} &  MF & MMSE denoiser & / \\
 \hline
EP/OAMP/VAMP \cite{EP}  &  LMMSE & MMSE denoiser & /   \\
 \hline
OAMP-Net \cite{MIMO_Detection20DL}   &  Learnable LMMSE & Divergence-free denoiser & Linear module  \\
\hline
GEPNet \cite{Kosasih20DL}  & LMMSE+GNN & MMSE denoiser & Linear module     \\
\hline
LDAMP\cite{DL2018HE}   & MF & CNN-based denoiser & Nonlinear module     \\
\hline
LDGEC\cite{DL2022HE}   & LMMSE & CNN-based denoiser & Nonlinear module     \\
\hline
\end{tabular}
\end{table*}

\section{Proposed AMP-GNN Network}\label{Sec:AMP_GNN}
In this section, we propose an AMP-GNN network for solving statistical inference problems in wireless communications. First, we illustrate the network structure of the proposed AMP-GNN, which is obtained by unfolding the AMP algorithm and incorporating an MPNN module. Then, the AMP algorithm and MPNN module are introduced in detail, respectively. Finally, several key properties of the proposed network and the reasons that lead to its better performance are identified.
\subsection{AMP-GNN Architecture}
\begin{figure*}
  \centering
  \includegraphics[width=16cm]{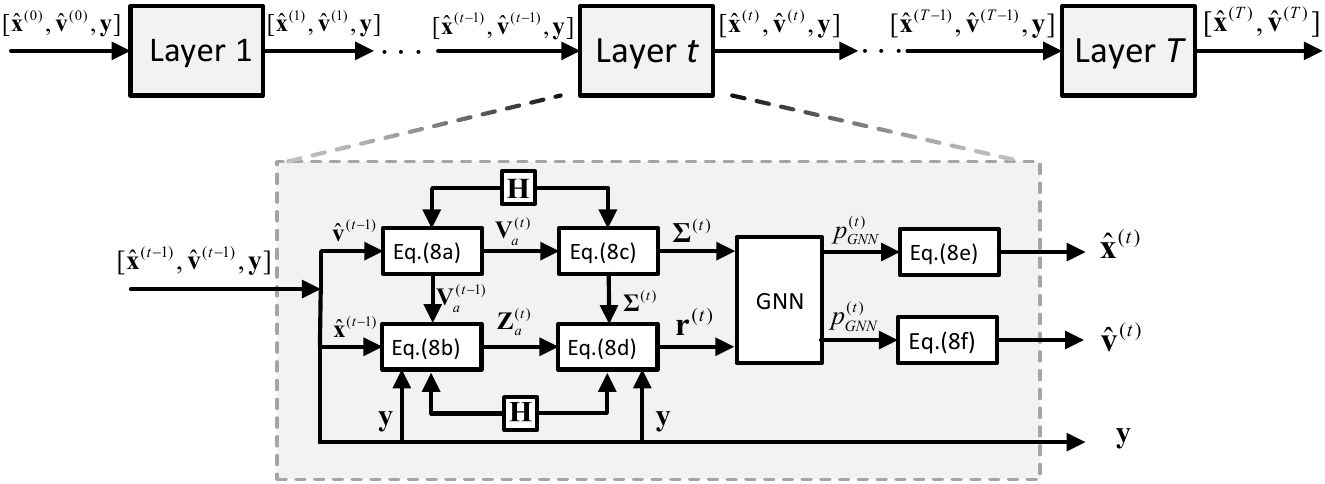}
  \caption{\small{.~~The structure of the proposed AMP-GNN network.}}\label{fig:AMPGNN}
\end{figure*}

GNNs have several important advantages, such as modeling interactions between pairs of nodes. We can also exploit them to mitigate the correlation in the decoupled AWGN channels for the AMP algorithm. The block diagram of the AMP-GNN is illustrated in Fig.\,\ref{fig:AMPGNN}. The network consists of $T$ cascade layers, and each layer has the same structure that contains a GNN module and the conventional AMP algorithm. The input of the AMP-GNN is the received signal $\mathbf{y}$, with the initial value setting as $\hat{\mathbf{x}}^{(1)}=\mathbf{0}$ and $\hat{\mathbf{v}}^{(1)}=\frac{N}{M}\mathbf{1}_N$, and the output is the final estimate $\hat{\mathbf{x}}^{(T)}$ of signal $\mathbf{x}$. For the $t$-th layer of the AMP-GNN, the inputs are the estimated signal $\hat{\mathbf{x}}^{(t-1)}$ and $ \hat{\mathbf{v}}^{(t-1)}$ from the $(t-1)$-th layer and the received signal $\mathbf{y}$. 
Finally, the AMP-GNN is executed iteratively until terminated by a fixed number of layers. 
In the next subsection, we introduce the AMP algorithm and the structure of the adopted MPNN module in detail.
\subsection{AMP algorithm}\label{sec:AMP}
\begin{algorithm}[!t]
\caption{AMP algorithm}
\label{Algorithm:BAMP}
{
\begingroup
\textbf{1. Input:} $\by$, $\bA$, $\sigma^2$, $\sfP(\bx)$. \\
\textbf{2. Initialization:} $\hat{x}_n^{(1)}=0$, $\hat{v}_n^{(1)}=\frac{N}{M}$, $Z_m^{(0)}=y_m$.\\
\textbf{3. Output:} $\hat{\bx}^{(T)}$.\\
\textbf{4. Iteration:} \\
\For{$t=1,\cdots, T$}
{
 \setlength\abovedisplayskip{0pt}
 \setlength\belowdisplayskip{0pt}
 \begin{subequations}
 \begin{align}
 V_m^{(t)}&=\sum_{n=1}^N |a_{mn}|^2\hat{v}_n^{(t)} \label{Equ:AMP1}\\
 Z_m^{(t)}&=\sum_{n=1}^Na_{mn}\hat{x}_n^{(t)}-\frac{V_m^{(t)}(y_m-Z_m^{(t-1)})}{\sigma^2+V_m^{(t-1)}} \label{Equ:AMP2}\\
 \Sigma_n^{(t)}&=\left(\sum_{m=1}^M\frac{|a_{mn}|^2}{\sigma^2+V_m^{(t)}}\right)^{-1} \label{Equ:AMP3}\\
 r_n^{(t)}&= \hat{x}_n^{(t)}+\Sigma_n^{(t)}\sum_{m=1}^M\frac{a_{mn}^{*}(y_m-Z_m^{(t)})}{\sigma^2+V_m^{(t)}} \label{Equ:AMP4}\\
 \hat{x}_{n}^{(t+1)}& =\mathbb{E}\{x_n|r_n^{(t)},\Sigma_n^{(t)}\}
 \label{Equ:AMP5}\\
 \hat{v}_{n}^{(t+1)}& =\text{Var}\{x_n|r_n^{(t)}, \Sigma_n^{(t)}\}
 \label{Equ:AMP6}
 \end{align}
 \end{subequations}
}
\endgroup
}
\end{algorithm}

The AMP algorithm was first proposed to solve sparse linear inverse problems in compressed sensing \cite{Donohol_AMP}, and has been widely used in various scenarios \cite{gaussian_mixture,AMPJSTSP,AMPcoded}. It admits a rigorous analysis based on state evolution. Such advantages motivated researchers to apply the AMP algorithm in wireless communications. In \textbf{Algorithm \,\ref{Algorithm:BAMP}}, we summarize the AMP algorithm for an arbitrary signal\footnote{Note that we consider a complex-valued AMP-based MIMO detector in  \textbf{Algorithm \,\ref{Algorithm:BAMP}} and the equivalent real-valued form can be derived with the equivalent real-valued representation accordingly.}, where $m$ and $n$ are the indexes of $\by$ and $\bx$, respectively. The main principle of the algorithm is to decouple the posterior probability $\sfP(\bx|\by,\bA)$ into a series of $\sfP(x_{n}|\by,\bA)$, for  $n=1,2,\ldots,N$, in an iterative way. In particular, $\sfP(x_{n}|\by,\bA)$ is assumed to be a Gaussian distribution that is obtained from the equivalent AWGN model in (\ref{eqAWGN}).
Equations (\ref{Equ:AMP5}) and (\ref{Equ:AMP6}) perform the posterior mean and variance estimation for the equivalent AWGN model (\ref{eqAWGN}) and exact expressions are related to the prior information of the signal. If  the transmitted symbol is assumed to be drawn from the $Q$-QAM set $\mathcal{S}=\{s_{1}, s_{2}, \ldots, s_{Q}\}$, the results in (\ref{Equ:AMP5}) and (\ref{Equ:AMP6}) are given by
\begin{equation}\label{eqmean}
  \hat{x}^{(t+1)}_{n} = \frac{\sum_{s_{i} \in \mathcal{S}}s_{i}\mathcal{N}_{\bbC}(s_{i};r_n^{(t)}, \Sigma_n^{(t)})p(s_{i})}{\sum_{s_{i} \in \mathcal{S}}\mathcal{N}_{\bbC}(s_{i};r_n^{(t)}, \Sigma_n^{(t)})p(s_{i})},
\end{equation}
\begin{equation}\label{eqvar}
  \hat{v}^{(t+1)}_{n} = \frac{\sum_{s_{i} \in \mathcal{S}}|s_{i}|^{2}\mathcal{N}_{\bbC}(s_{i};r_n^{(t)}, \Sigma_n^{(t)})p(s_{i})}{\sum_{s_{i} \in \mathcal{S}}\mathcal{N}_{\bbC}(s_{i};r_n^{(t)}, \Sigma_n^{(t)})p(s_{i})}
  -|\hat{x}^{(t+1)}_{n}|^{2}.
\end{equation}

As can be observed in \textbf{Algorithm \,\ref{Algorithm:BAMP}}, the performance of the AMP algorithm is mainly determined by the accuracy of the equivalent AWGN model (\ref{eqAWGN}).  In \cite{dynamicAMP}, it was shown that the equivalent AWGN model is asymptotically accurate when the dimensions of the system tend to infinity, i.e.,  $M, N \rightarrow \infty$. However, in practical finite-dimensional systems, the performance of the AMP algorithms is far from optimal and even has an error floor owing to the inaccurate assumption, which motivates us to improve the AMP algorithm with the advanced DL technique, i.e., GNNs.
\subsection{MPNN Module}
\begin{figure*}
  \centering
  \includegraphics[width=16cm]{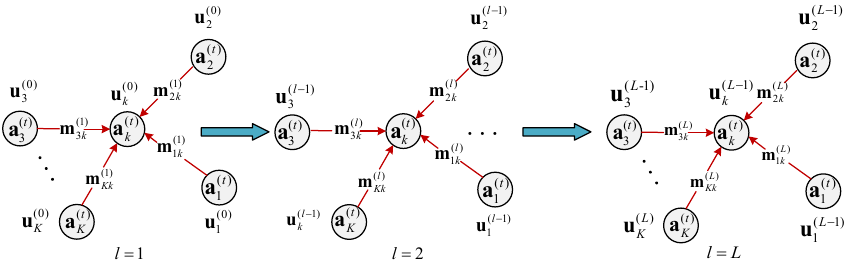}
  \caption{.~~The block diagram of the structure for MPNN.}\label{fig:GNN_structure}
\end{figure*}

GNNs have been recently adopted for  wireless communications as they can incorporate the graph topology of the wireless network into the neural network design and optimize the objectives with the data-driven methods \cite{YifeiGNN2}. In particular, the antennas or users are considered  as the nodes while the channels are considered as the edges to construct the GNNs. 
In our proposed AMP-GNN framework, we adopt the MPNN \cite{Gilmer17}. The MPNN, as illustrated in Fig. \,\ref{fig:GNN_structure}, consists of $L$ cascade layers, where each node $n$ is connected to all other nodes. The aggregation module, comprising  GRU module $\mathsf{U}$ and a linear network, is utilized by each node to update the node hidden vector $\mathbf{u}_n^{(l)}$. Note that each node $n$ shares the same weights of the GRU module $\mathsf{U}$ and linear network. Furthermore,  a message $m_{jn}^{(l)}$  must be calculated for each pair of variable nodes $n$ and $j$ during the $l$-th round of the MPNN. Specifically, the message $m_{jn}^{(l)}$ is obtained using the multi-layer perceptron (MLP) module $\mathsf{D}$, with each pair $n$ and $j$ sharing the same weights. The reason to adopt the  MPNN is that it can unify various GNNs and graph convolutional network approaches \cite{GNN_Survey}. Furthermore, the propagation and aggregation modules in MPNN are very similar to the message passing operation on the factor graph. As a result, its structure is very suitable for  statistical inference problems. As mentioned in Section \ref{sec:AMP}, the AMP algorithm can decouple the linear mixing model (\ref{eqfirst}) into the equivalent AWGN model (\ref{eqAWGN}). However, the decoupling is not accurate enough which means that the $N$ equivalent AWGN channels are not independent; i.e.,  there exists structured dependency among these nodes\footnote{As demonstrated in \cite{Donohol_AMP}, if the asymptotic conditions of AMP are satisfied, the structured dependency of the nodes is eliminated due to the central limit theorem (CLT). However, this conclusion is highly dependent on the large-system system limit. In finite-dimensional systems, the structured dependency of nodes still exists. Additionally, for high-order modulation symbols (e.g.,16-QAM and 64-QAM), there is an SER floor when $M/N > \beta_{th}$, indicating that the structured dependency of nodes cannot be ignored in these cases\cite{Jeon_AMP}.}. This motivates us to adopt MPNN to exploit the correlation between the equivalent AWGN channels and mitigate interference.


In the literature of  machine learning, the pair-wise MRF has been utilized to model the structured dependency of a set of random variables $\bx = \{x_{1},\cdots,x_{N}\}$ by an undirected graph $G = \{V, E\}$. It can be adopted to model the correlation of the nodes in GNNs. Specifically, the $n$-th variable node is characterized by a self potential $\phi(x_n)$, and the $(n,j)$-th pair of the edge is characterized by a pair potential $\psi(x_n,x_j)$, which are given by
\begin{subequations}\label{phi_and_psi}
            \begin{equation}\label{phi}
\phi(x_n) = \sfee \sfx \sfp \left( \frac{1}{\sigma^2} \by^T \ba_n x_n - \frac{1}{2}  \ba_n^T \ba_n x_n^2  \right) \sfP(x_n),
            \end{equation}
            \begin{equation}\label{psi}
\psi(x_n,x_j) = \sfee \sfx \sfp \left(- \frac{1}{\sigma^2} \ba_n^T \ba_j x_n x_j\right), n,j  \in 1,\ldots, N, \quad \mathrm{and} \quad n \neq j
            \end{equation}
\end{subequations}
respectively, where $\ba_n$ denotes the $n$-th column of the matrix $\bA$. The posterior probability $ p_{\mathrm{GNN}}(\bx|\by)$ corresponding to the pair-wise MRF for the statistical inference problem  can be obtained by the MPNN and written as \cite{Scotti20GNN}
\begin{equation}
 p_{\mathrm{GNN}}(\bx|\by) 
 =\frac{1}{Z} \prod_{n=1}^N \phi(x_n) 
 \prod_{\substack{j=1\\j \neq n}}^N \psi(x_n,x_j),
 \label{p_x_y_GNN}
\end{equation}
where $Z$ is a normalization constant. In particular,  $\phi(x_n)$ and $\psi(x_n,x_j)$ can be represented by the information of the nodes and edges in GNNs, respectively. Equation \eqref{p_x_y_GNN} implies that a well-trained MPNN can characterize the posterior probability $ p_{\mathrm{GNN}}(\bx|\by)$, which is the key of statistical inference.

The MPNN adopted in the AMP-GNN is composed of three main modules: a \emph{propagation module}, an \emph{aggregation module}, and a \emph{readout module}. The first two modules operate in all layers while the readout module is involved only after the last layer.  To better understand the structure of the MPNN, we first elaborate on the following concepts about the GNNs. In particular, we mainly introduce the definitions of node and edge and their associated attributes.

\begin{itemize}
  \item \textbf{Node:} In GNNs, each node $n \in V$ represents the $n$-th user or $n$-th antenna in the wireless systems.
  \item \textbf{Node Attribute:} Each node has an assigned node attribute $\ba_{n}$ that is constant when exchanging information between different nodes. In the proposed AMP-GNN, the MPNN in the $t$-layer takes the output from the linear module of the AMP algorithm as a node attribute.
  \item \textbf{Edge:} An edge  $e_{n,j} \in E $ is to connect node $n \in V$  and $j \in V$. Whether the edge exists or not depends on the graph structure for the target problem.
  \item \textbf{Edge Attributes:} Each edge $e_{n,j}$ has an assigned edge attribute $\mathbf{f}_{jn}$ that is constant when computing the message. In the proposed AMP-GNN, the MPNN uses the CSI and noise level as the edge attributes.
  \item \textbf{Hidden vector:} Each node $n$ has a hidden vector $\bu_{n}$ updated in  different rounds of the MPNN, and will be used to compute the output of the GNN.
  \item \textbf{Message:} The incoming messages $\bm_{jn}$ from its connected edges are utilized to update the node feature vector $\bu_{n}$.
\end{itemize}

When designing GNNs, we need to first define the node and edge attributes. As the MPNN in the $t$-layer of the AMP-GNN takes the output from the linear module in the AMP algorithm as the input, it is natural to  incorporate the mean $r_{n}^{(t)}$ and variance $\Sigma_n^{(t)}$  obtained from  (\ref{Equ:AMP3}) and (\ref{Equ:AMP4}) into the attribute $\bd_n^{(t)}$  of the variable node $x_n$ by concatenating the mean and variance as
\begin{equation}\label{concat}
\bd_n^{(t)} = \left[ r_{n}^{(t)}, \Sigma_n^{(t)} \right].
\end{equation}

The second step is to define the initialized hidden vector $\bu_n^{(l)}$ for each node $x_{n}$.  We consider the initial value calculated from encoding the information of the received signal $\by$, corresponding linear vector $\bd_n^{(t)}$, and noise variance $\sigma^2$.  The encoding process is implemented by using a single layer neural network given by
\begin{equation}\label{GNN_init}
 \bu_n^{(0)} = \bW_1 \cdot [\by^T \ba_n, \ba_n^T \ba_n, \sigma^2]^T + \bb_1,
\end{equation}
where $\bW_1 \in \mathbb{R}^{N_u \times 3}$ is a learnable matrix, $\bb_1 \in \mathbb{R}^{N_u}$ is a learnable vector, and $N_u$ is the size of the hidden vector.
The edge attribute  $\mathbf{f}_{jn} \triangleq \left[ \ba_n^T \ba_j, \sigma^2 \right] $ is obtained by extracting the pair potential information from \eqref{psi} and is then utilized for the message passing of the MPNN. 
Based on these definitions and operations, we elaborate on the details of each module in the MPNN in the following, including the propagation, aggregation, and readout modules.

\subsubsection{Propagation module}
For any pair of variable nodes $x_n$ and $x_j$ in  (\ref{eqAWGN}), we assume that there is an edge to connect them. In the $l$-th round of the MPNN, each edge first concatenates the hidden vectors $\bu_n^{(l-1)}$ and $\bu_j^{(l-1)}$ with its own edge attribute $\mathbf{f}_{jn}$ as
\begin{equation}\label{eqcnl}
\bc_{n}^{(l-1)} = [\bu_n^{{(}l-1{)}}, \bu_j^{{(}l-1{)}}, \mathbf{f}_{jn}].
\end{equation}
Then, it uses the concatenated features $\bc_{n}^{(l-1)}$ as the input for the multi-layer perceptron (MLP). Therefore, the output of the MLP is given by
\begin{equation}\label{factor_to_var}
\bm_{jn}^{(l)} = {\sfD} \left( \bc_{n}^{(l-1)} \right),
\end{equation}
where  $\sfD$ is the MLP operation. In the propagation module, each edge has an MLP with two hidden layers of sizes $N_{h_1}$ and $N_{h_2}$ and an output layer of size $N_u$.  
Furthermore, the rectifier linear unit (ReLU) activation function is used at the output of each hidden layer. Finally, the outputs $\bm_{jn}^{(l)}$ are fed back to the nodes as shown in Fig.\,\ref{fig:GNN_structure}. In particular, $\bm_{jn}^{(l)}$ can be interpreted as the message transmitted from node $j$ to node $n$.

\subsubsection{Aggregation  module}
The $n$-th variable node  sums all the incoming messages $\bm_{jn}^{(l)}$ from its connected edges  and concatenates the sum of the $\bm_{jn}^{(l)}$  with the $t$-layer node attribute $\bd_n^{(t)}$ as $\bm_n^{(l)}  = \left[\sum_{\substack{j=1\\j \neq n}}^N \bm_{jn}^{(l)},  \bd_n^{(t)} \right]$. 
Then, the message $\bm_n^{(l)}$ is used to compute the node hidden  vector $\bu_n^{(l)}$ as
 \begin{subequations}\label{GRU_MLP}
            \begin{equation}\label{GRU}
 \bg_n^{(l)} = {\sfU} \left( \bg_n^{(l-1)}, \bm_n^{(l)}  \right),
            \end{equation}
            \begin{equation}\label{MLP}
 \bu_n^{(l)}= \bW_2 \cdot \bg_n^{(l)} + \bb_2,
            \end{equation}
\end{subequations}
where the function $\sfU$ is specified by the gated recurrent unit (GRU) network, whose current and previous hidden states are $\bg_n^{(l)} \in \mathbb{R}^{N_{h_1} }$ and  $\bg_n^{({l}-1)} \in \mathbb{R}^{N_{h_1} }$, respectively.  $\bg_n^{(l)}$ can be interpreted as the intermediate variables to update the
hidden  vector $\bu_n^{(l)}$. In (\ref{MLP}), $\bW_2 \in \mathbb{R}^{N_u \times N_{h_1}}$ is a learnable matrix, and  $\bb_2\in \mathbb{R}^{N_u}$ is a learnable vector. The updated feature vector $\bu_n^{(l)}$ is then sent to the propagation module for the next iteration.

\subsubsection{Readout module}
After $L$ rounds of the message passing between the propagation and aggregation modules, a readout module ${\sfR}$ is utilized to output the estimated result. We need to especially design the final results from the MPNN for the next AMP-GNN iteration. Here we take the classification problem as an example. Specifically, we consider a readout module that is utilized in each node to output the final estimated distribution $\sfp_{\mathrm{GNN}}^{(t)}(x_n=s_{i}|\by)$ for the $t$-layer of the AMP-GNN and is given by\footnote{Here we assume the MPNN is utilized for  classification problems, thus the output of the Readout module is a discrete distribution $\sfp_{\mathrm{GNN}}^{(t)}(x_n=s_{i}|\by)$. We can also obtain the continuous value of the output for  regression problems. }
\label{Readout}
\begin{equation}\label{Readout1}
\sfp^{(t)}_{ \mathrm{GNN}}(x_n|\by)  = {\sfR} \left(  \bu_n^{(L)} \right),
\end{equation}

The readout function ${\sfR} $ consists of an MLP with two hidden layers of sizes $N_{h_1}$ and $N_{h_2}$, and ReLU activation is utilized at the output of each hidden layer. Finally, the hidden and node hidden vectors are updated as
\begin{equation}\label{GNN_reset_index}
 \bg_n^{(0)}  \leftarrow   \bg_n^{(L)} \text { and }\bu_n^{(0)} \leftarrow   \bu_n^{(L)}, \quad n= 1, \dots,N,
\end{equation}
for the MPNN initialization in the next AMP-GNN iteration. The obtained distribution $\sfp^{(t)}_{ \mathrm{GNN}}(x_n|\by)$ will be passed to the nonlinear module in the AMP to further refine the estimated result.

To better understand the output of the MPNN, we take the massive MIMO detection as an example. As the  transmitted signal is drawn from the discrete $Q$-QAM set, we further refine the $\sfp_{\mathrm{GNN}}^{(t)}(x_n=s_{i}|\by)$ with prior information $\sfP(x_n)$ and compute the posterior mean and variance for the next layer of the AMP-GNN, which are given by
\begin{subequations}\label{pyGNN}
            \begin{equation}\label{pyGNN1_1}
 \hat{x}_{n}^{(t+1)}=\mathbb{E}\{x_n|\sfp^{(t)}_{ \mathrm{GNN}}(x_n|\by)\},
            \end{equation}
            \begin{equation}\label{pyGNN2_2}
 \hat{v}_{n}^{(t+1)}=\text{Var}\{x_n|\sfp^{(t)}_{ \mathrm{GNN}}(x_n|\by)\}.
            \end{equation}
\end{subequations}
After computing (\ref{pyGNN1_1}) and (\ref{pyGNN2_2}), the posterior mean $\hat{x}_{n}^{(t+1)}$ and $\hat{v}_{n}^{(t+1)}$ are used for the next AMP-GNN iteration. Finally, the AMP-GNN is executed iteratively until terminated by a fixed number of layers. Note that the expectation and variance in (\ref{pyGNN1_1}) and (\ref{pyGNN2_2}) are computed with respect to $\sfp_{\mathrm{GNN}}^{(t)}(x_n|\by)$. This is the main difference between AMP and AMP-GNN. In particular, the $\sfp_{\mathrm{GNN}}^{(t)}(x_n=s_{i}|\by)$ is assumed to be Gaussian pdf and obtained by the equivalent AWGN model. By contrast, $\sfp_{\mathrm{GNN}}^{(t)}(x_n=s_{i}|\by)$ is obtained by learning from the data and is not the Gaussian pdf in the AMP-GNN. Thus, the inaccurate Gaussian pdf is refined by the GNNs.

\subsection{Properties of AMP-GNN}
The AMP-GNN enjoys several properties that are favorable to solve statistical inference problems in wireless communications, including permutation equivariance and
generalization to different numbers of users. These advantages enable the AMP-GNN to learn more efficiently, avoid over-fitting, and develop strong generalizability.
\subsubsection{Permutation Equivariance}
The first characteristic is  permutation equivariance. Consider a set $\mathcal{F}$ of all functions $f:\bbR^{M \times M}\rightarrow \bbR^{M}$  and a generic permutation matrix $\boldsymbol{\Pi}\in\{0,1\}^{M \times M}$, we have the following definition of  permutation equivariance.

\textbf{Definition 1.} A function $f \in \mathcal{F}$ is permutation equivariant if $\boldsymbol{\Pi}^{T}f(\bA,\by) = f(\bA\boldsymbol{\Pi},\by) $ for all matrices and all permutations $\boldsymbol{\Pi}$. In particular, if we permute the labels of the users or antennas in our network before computing the permutation equivariant function $f$, the individual output values are not changed but only permuted by this same permutation. Note that this is especially critical for our problem because the index of the node is arbitrary and should not play any role in the estimated result.

\textbf{Proposition 1.}
Assuming that $\boldsymbol{\Pi}$ is a permutation matrix, $\tilde{\bA} = \bA\boldsymbol{\Pi}$ denotes the permuted channel matrix, and $\tilde{\bx} = \boldsymbol{\Pi}^{T}\bx$ is the permutated signal vector. We have following permutated $\tilde{\bA}$ and $\tilde{\bx}$
\begin{equation}\label{eqvariant}
\by = \tilde{\bA}\tilde{\bx}+\bn = \bA\boldsymbol{\Pi}\boldsymbol{\Pi}^{T}\bx +\bn = \bA\bx +\bn.
\end{equation}
Thus, we have

\begin{equation}\label{eqvariant2}
\boldsymbol{\Pi}^{T} f_{\theta}(\bA, \by) = f_{\theta}(\bA\boldsymbol{\Pi}, \by),
\end{equation}
for the proposed AMP-GNN.

\emph{Proof:} Refer to Appendix A.

Permutation equivariance implies that the ordering of the users will not affect network performance. This is  because reordering the users simply permutes the columns of the matrix $\bA$ and is associated with an appropriate permutation of the symbol vector $\bx$. 
Furthermore, it also reduces the training sample complexity and training time compared to conventional MLPs and CNNs.  This is because for each training sample,
all its permutations are naturally contained in the training set.  But for MLPs and CNNs, data argumentation is required to achieve the same performance. Thus,
permutation equivariance is extremely helpful in reducing the training sample complexity and time compared to MLPs and CNNs.
\subsubsection{Generalize to Different Numbers of Users}
Most of the works on DL-based physical layer processing are trained and tested with a fixed number of antennas. However, the number of users (antennas) in practical massive MIMO systems may quickly change. For example, 
the number of active users is continuously changing with the dynamic nature of wireless networks. Training multiple networks with each network targeted to a specific number of users is not practical. In essence, constructing a network that can handle a varying number of users requires the network to be modular while sharing the same set of parameters for each user. In the AMP-GNN, the AMP algorithm is unrelated to the number of nodes and the dimension of the adopted MLP in MPNN is invariant with the number of users. Furthermore, they share the same parameters for different edges and users. As a result,  the proposed AMP-GNN has the ability to handle a varying number of users with a single model 
and we can train the AMP-GNN with a specific dimension and apply them to different settings.
\subsection{Why AMP-GNN can Enhance AMP?}
In Section \ref{sec:Bayesian}, we have elaborated on the derivation of the BP algorithm for Bayesian inference. However, the message updates in (\ref{Chapter_BP}) involve a high dimensional integral and true pdf, which have extremely high complexity. To this end, the AMP algorithm is derived by exploiting the central limit theorem and Taylor expansion to simplify the BP algorithm. On the other hand, the AMP algorithm can be derived alternatively from the perspective of EP \cite{Meng_EP}. In particular, the BP update in (\ref{Chapter_BP}) can be approximated by
\begin{align}\label{Chapter_AMP1}
\mu_{n\rightarrow m}^{(t+1)}(x_n)&\propto  \frac{\mathrm{Proj}[\sfP(x_n)\prod_{b} \mu_{b \rightarrow n}^{(t)}(x_n)]}{\mu_{m \rightarrow n}^{(t)}(x_n)},\\
\mu_{m\rightarrow n}^{(t)}(x_n)&\propto \int \frac{\mathrm{Proj}[\sfP(y_m|\bx)\prod_{j\ne i}^N\mu_{j\rightarrow m}^{(t)}(x_j)\text{d}\bx_{\backslash n}]}{\mu_{n \rightarrow m}^{(t)}(x_n)},
\end{align}
where $\mathrm{Proj}[\sfp]$ is  the projection of a distribution $\sfp$ to a distribution set $\mathcal{F}$ defined as
\begin{equation}\label{eqF}
  \mathrm{Proj}_{\cF}[\sfp] = \arg \min_{q\in\cF} D(\sfp\|\sfq),
\end{equation}
and $D(\sfp\|\sfq)$ denotes the Kullback-Leibler divergence. To reduce the overhead of transmitting the message, $\mathcal{F}$ is assumed to be the Gaussian distribution. This is because the Gaussian distribution can be fully characterized by its mean and variance, and thus only the mean and variance need to be calculated and passed. First,  we assume that $\mu_{n\rightarrow m}^{(t)}(x_n)\propto \cN_{\bbC}(x_{j};\hat{x}_{j\rightarrow m}^{t},v_{j\rightarrow m}^{t})$, and therefore $\prod_{b} \mu_{b \rightarrow n}^{(t)}(x_n)$ is the product of Gaussian distributions. Consider $y_{m} = a_{mn}x_{n}+\sum_{j\neq n}a_{mj}x_{j}+n_{m}$ and define $Z_{mn} = \sum_{j\neq n}|a_{mn}|^{2}v_{j\rightarrow m}^{t}\sim  \cN_{\bbC}(x;Z_{m\rightarrow n}^{t},V_{m\rightarrow n}^{t})$, we have
\begin{subequations}\label{eq:mja}
\begin{align}
Z_{m \rightarrow n}^{t} = \sum_{j \neq n}a_{mj}x_{j\rightarrow m}^{t}, \quad \mathrm{and} \quad
V_{m \rightarrow n}^{t} = \sum_{j \neq n}|a_{mj}|^{2}v_{j\rightarrow m}^{t}.
\end{align}
\end{subequations}
Then, we obtain  $\int [\sfP(y_m|\bx)\prod_{j\ne i}^N\mu_{j\rightarrow m}^{(t)}(x_j)\text{d}\bx_{\backslash n}]  \propto  \cN_{\bbC}\bigg( x_{n};  \frac{y_{m}-Z_{m \rightarrow n}^{t}}{a_{mn}}, \frac{\sigma^{2}+V_{m \rightarrow n}^{t}}{|a_{mn}|^{2}}\bigg)$. As a result, we have  $\mu_{m\rightarrow n}^{(t)}(x_n) \propto
\cN_{\bbC}\bigg(x_{n}; \frac{y_{m}-Z_{m \rightarrow n}^{t}}{a_{mn}}, \frac{\sigma^{2}+V_{m \rightarrow n}^{t}}{|a_{mn}|^{2}}\bigg)$. By adopting the Gaussian product lemma\footnote{In the derivation, we use the result that the product of multiple Gaussian distributed random variables yields another Gaussian random variable \cite{Gaussian_ML}. Take the product of two Gaussian as an example, we have $\mathcal{N}(x|a,A)\mathcal{N}(x|b,B)=Z\mathcal{N}(x|c,C)$, where $c=C(A^{-1}a+B^{-1}b)$, $C=(A^{-1}+B^{-1})^{-1}$, and
$Z=\frac{1}{\sqrt{2\pi(A+B)}}\exp\left(-\frac{(a-b)^2}{2(A+B)}\right)$.}, we have $\prod_{b}\mu_{b \rightarrow n}^{(t)}(x_n) \propto  \cN_{\bbC}(x_{n};r_{n}^{t},\Sigma_{n}^{t})$, where
\begin{subequations}\label{eqmba}
\begin{align}
\Sigma_{n}^{t} = \bigg(\sum_{m}\frac{|a_{mn}|^{2}}{\sigma^{2}+V_{m \rightarrow n}^{t}}\bigg)^{-1}, \\
\mu_{n}^{t} = \Sigma_{n}^{t}\bigg(\sum_{m}\frac{h_{mn}^{*}(y_{m}-Z_{m \rightarrow n}^{t})}{\sigma^{2}+V_{m \rightarrow n}^{t}}\bigg).
\end{align}
\end{subequations}
According to (\ref{Chapter_AMP1}), we have $\mathrm{Proj}[\sfP(x_{n})\prod_{b}\mu_{b \rightarrow n}^{(t)}(x_n)] \propto \cN_{\bbC}(x_{n}; \hat{x}_{n}^{t+1},\hat{v}_{n}^{t+1})$ and
\begin{subequations}\label{ex}
\begin{align}
 \hat{x}_{n}^{(t+1)}& =\mathbb{E}\{x_n|r_n^{(t)},\Sigma_n^{(t)}\},  \\
 \hat{v}_{n}^{(t+1)}& =\text{Var}\{x_n|r_n^{(t)}, \Sigma_n^{(t)}\}.
\end{align}
\end{subequations}
As the $\mathrm{Proj}$ operation utilizes a Gaussian distribution to approximate $\mathrm{Proj}[\sfP(x_{n})\prod_{b}\mu_{b \rightarrow n}^{(t)}(x_n)]$, its accuracy  depends on the prior information $\sfP(x_{n})$ and the approximated message $\prod_{b} \mu_{b \rightarrow n}^{(t)}(x_n)$. However, the approximated message $\prod_{b} \mu_{b \rightarrow n}^{(t)}(x_n)$ is far from the Gaussian pdf and  $\sfP(x_{n})$ is not exactly known for some applications in practical massive MIMO systems \footnote{The classical AMP algorithm proved Gaussianity based on the GLT in the large-system limit \cite{Donohol_AMP}. However, in practical finite-dimensional MIMO systems, Gaussianity cannot be guaranteed. As emphasized in \cite{Jeon_AMP}, the statistics of $\mathbf{r}^{(t)}$ are not Gaussian and therefore cannot be accurately tracked by state evolution, which incurs the performance loss. Additionally, as analyzed in \cite{Jeon_AMP}, there are parameter regimes where the AMP-based MIMO detection only achieves suboptimal performance even in the large-system limit. In the cases where the performance of the AMP is far from optimal, GNN can be used for performance enhancement.}. To this end, GNNs can learn an accurate $\mathrm{Proj}$ from the data beyond the specific distribution $\sfP(x_{n})$ and inaccurate approximation $\prod_{b} \mu_{b \rightarrow n}^{(t)}(x_n)$ by its powerful capability, which is the underlying reason why the AMP-GNN outperforms the AMP algorithm. In other words,  GNNs can learn an accurate approximation for the Kullback-Leibler divergence.
\begin{table*}
\centering
	\caption{Computational complexity (the number of multiplications) of different detectors. 
}
	\label{tab:complexity}
    \begin{tabular}{@{}lcccccc@{}}
    \toprule
    \diagbox{MIMO settings}{Detectors}&OAMP&GNN&GEPNet&AMP&AMP-GNN&EP\\
    \midrule
    \midrule
    $64 \times 64$ &$8.22\times10^{6}$&$1.17\times10^{6}$&$5.11\times10^{6}$&$1.78\times10^{5}$&$2.35\times10^{6}$& $2.93\times10^{6}$ \\
    \midrule
    $256 \times 256$ &$5.21\times10^{8}$&$1.27\times10^{7}$&$2.02\times10^{9}$&$2.68\times10^{6}$&$1.93\times10^{7}$& $1.85\times10^{9}$ \\
    \midrule
    $1024 \times 1024$ &$3.33\times10^{10}$&$5.56\times10^{8}$&$1.24\times10^{10}$&$4.22\times10^{7}$&$6.14\times10^{8}$& $1.18\times10^{10}$ \\
    \bottomrule
    \end{tabular}
\end{table*}
\section{Application to Massive/Ultra-Massive MIMO Systems}\label{Applications}
As mentioned in Section \ref{Sec:inference}, abundant applications in wireless communications can be categorized into statistical inference problems. Although we have provided a general framework, some dedicated network design should be considered for specific applications. To show the effectiveness of the AMP-GNN framework, we take the massive/ultra-massive MIMO detection as an example in this section. We first elaborate the application of AMP-GNN for massive/ultra-massive MIMO detection. Then, the computational complexity of the proposed AMP-GNN-based MIMO detector is  analyzed.
\subsection{Massive/Ultra-Massive MIMO Detection}
In massive and ultra-massive MIMO systems, the dimension of  antenna arrays is extremely large. Efficient MIMO detection algorithms, which balance performance and complexity, are of significant importance to fully unleash the  potential of such  large-scale systems. We consider an uplink multi-user MIMO (MU-MIMO) systems where the base station (BS) equipped with $M$ antennas serves $N$ single-antenna users. Assuming that the symbol vector $\bx\in\mathbb{C}^{M\times1}$ is transmitted over a Rayleigh fading channel $\bH\in \mathbb{C}^{M \times N}$ and each element of $\bH$ and $\bx$ is drawn from an independent and identically distributed (i.i.d.) complex Gaussian distribution and a $Q$-QAM constellation, respectively. Thus, the received signal $\by\in\mathbb{C}^{M\times1}$ is given by
$\by=\bH\bx+\bn.$

One challenge of applying the AMP-GNN framework to MIMO detection is how to design the readout module. As the MIMO  detection is a classification problem,
the output size of  ${\sfR} $ is the cardinality of the real-valued constellation set, i.e., $\sqrt{Q}$. We further use the softmax function to restrict the output of each node in a probabilistic form,
\begin{equation}\label{Readout2}
{\tilde \sfp}_{\mathrm{GNN}}^{(t)}(x_n=s_{i}|\by)  = \frac{{\sfee \sfx \sfp } \left( \sfp^{(t)}_{G\mathrm{NN}}(x_n=s_{i}|\by) \right)}  {\sum_{s_{i}\in \mathcal{S}}  {\sfee \sfx \sfp} \left(\sfp^{(t)}_{ \mathrm{GNN}}(x_n=s_{i}|\by) \right)}, \quad s_{i}\in \mathcal{S}.
\end{equation}

As the distribution of the transmitted signal is known (i.e., $Q$-QAM), we further refine the $\sfp_{\mathrm{GNN}}^{(t)}(x_n=s_{i}|\by)$ with prior information $\sfP(x_n)$ and compute the posterior mean and variance for the next layer of the AMP-GNN, which are given by
\begin{subequations}\label{pyGNN}
            \begin{equation}\label{pyGNN1}
 \hat{x}_{n}^{(t+1)}=\mathbb{E}\{x_n|{\tilde \sfp}_{\mathrm{GNN}}^{(t)}(x_n=s_{i}|\by)\},
            \end{equation}
            \begin{equation}\label{pyGNN2}
 \hat{v}_{n}^{(t+1)}=\text{Var}\{x_n|{\tilde \sfp}_{\mathrm{GNN}}^{(t)}(x_n=s_{i}|\by)\},
            \end{equation}
\end{subequations}
where the expectation and variance are computed with respect to ${\tilde \sfp}_{\mathrm{GNN}}^{(t)}(x_n=s_{i}|\by)$.  
\subsection{Complexity Analysis}
In this section, we analyze the computational complexity of the AMP-GNN-based MIMO detector and compare it with existing DL and message passing based MIMO detectors. Specifically, the complexity of the AMP detector is  $\mathcal{O}(MN)$ due to the matrix-vector multiplication while the complexity for MPNN is  $\mathcal{O}(NN_{h1}N_{h2})$ which accounts for the MLP operation. Therefore, the computational complexity of the AMP-GNN is $\mathcal{O}(MN+NN_{h1}N_{h2})$, dominated by the complexity of the AMP and MPNN. In contrast, the complexity of the GEPNet is $\mathcal{O}(M{N}^2+NN_{h1}N_{h2})$ which includes the computational complexity of the EP and GNN.

To conduct a fair comparison of the computational complexity, we will use the number of multiplications as the metric and show the exact values for different MIMO settings with quadrature phase shift keying (QPSK) symbols in Table \,\ref{tab:complexity}. Compared with the state-of-the-art DL-based MIMO detectors, i.e., GEPNet, the AMP-GNN entails a much lower complexity. In particular, the ratio between the complexity of the AMP-GNN and GEPNet is dramatically reduced when the number of users increases. For example, the ratio between the complexity of the AMP-GNN and GEPNet is only $45.99\%$ when $M=N=64$  while the ratio is significantly reduced to $4.95\%$ when $M=N=1024$.  This is because the complexity of matrix inversion  in the GEPNet is the dominant term, which is prohibitively high when the number of antennas and user is large. In contrast, the AMP-GNN only involves  matrix-vector multiplications, which is a favorable feature for future ultra-massive MIMO systems. On the other hand, the propagation and aggregation modules on each node and edge are executed in parallel, which means the MPNN can be further processed in a distributed manner and the time complexity can also be reduced. This is also a great advantage for distributed ultra-massive MIMO systems.
\begin{table}[t]
	\centering
\caption{Simulation parameters of the AMP-GNN for massive MIMO detection.}
\begin{tabular}{|c|c|}
  \hline
  \hline
  Simulation parameters & Value \\
  \hline
  Number of users ($N$) & 16, 24, 32,  64 \\
  \hline
  Number of antennas ($M$)  & 16, 24, 32,  64\\
  \hline
  Number of realizations (d) & 100000 \\
  \hline
  The hyperparameters for MPNN & $N_{h_1}=16$,  $N_{h_2}=8$,  and $N_u = 8$ \\
  \hline
  Modulation symbols ($Q$-QAM) & 4-QAM, 16-QAM, 64-QAM \\
   \hline
  Training SNR   & SNR = 20 dB \\
  \hline
\end{tabular}
\label{Table:simulation results}
\end{table}
\section{Simulation Results}\label{Simulation}
In this section, we mainly provide simulation  results of the AMP-GNN for MIMO detection and compare them with other MIMO detectors. We use the symbol error rate (SER) as the performance metric in our simulations. The signal-to-noise (SNR) of the system is defined as $\mathrm{SNR}=\frac{\mathbb{E}\|\mathbf{H}\mathbf{x}\|^{2}_{2}}{\mathbb{E}\|\mathbf{n}\|^{2}_{2}}$. To illustrate the effectiveness of our proposed AMP-GNN, we adopt several well-established MIMO detectors as baselines:
\begin{itemize}
  \item \textbf{MMSE}: A classical linear receiver for MIMO detection which inverts the received signal by applying the channel-noise regularized pseudo-inverse
        of the channel matrix.
  \item \textbf{AMP}: An efficient message passing algorithm for MIMO detection given in \textbf{Algorithm} \ref{Algorithm:BAMP} and implemented with $10$ iterations\footnote{It was found that a further increase in the number of iterations only offers a negligible performance gain. We set the same number of layers in other DL-based baseline methods for fair comparison.}.
  \item \textbf{OAMP-Net}: The OAMP-based model-driven DL detector developed in \cite{MIMO_Detection20DL}. Each layer requires computing a matrix pseudo-inverse and has 2 learnable parameters.
  \item \textbf{EP}: The EP-based MIMO detector with 10 iterations as proposed in \cite{EPdetector}.
  \item \textbf{GEPNet}: The GNN-enhanced EP detector proposed in \cite{Kosasih20DL} with $10$ layers.
\end{itemize}

\subsection{Implementation Details}
In the simulation, the AMP-GNN is implemented on the PyTorch platform. The number of layers of the AMP-GNN detector  is set to $T=10$ while the number of layers of the GNN is set to $L=2$. The training data consists of a number of randomly generated pairs $(\mathbf{x},\mathbf{y})$. The data $\mathbf{x}$ is generated from QAM modulation symbols. We train the network for 500 epochs with the same training and validation sets in each epoch. The training set contains 100,000 samples while the validation set contains 5,000 samples. The AMP-Net is trained using the stochastic gradient descent method and Adam optimizer. The learning rate is set to $0.001$ and the batch size is set to $64$.
We choose $L_{2}$ loss as the cost function, which is defined by,
\begin{equation}\label{eqloss}
  L_{2}(\mathbf{x},\hat{\mathbf{x}}^{(T)})=\|\mathbf{x}-\hat{\mathbf{x}}^{(T)}\|^{2}.
\end{equation}
\begin{figure}[h]
\begin{minipage}{3in}
  \centerline{\includegraphics[width=3.2in]{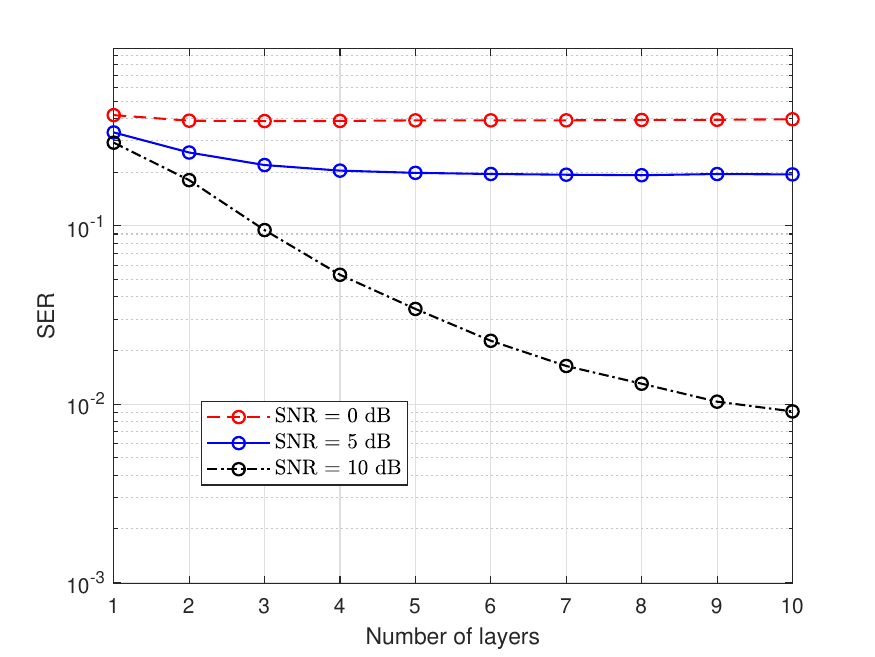}}
  \centerline{(a) 16-QAM.}
\end{minipage}
\hfill
\begin{minipage}{3in}
  \centerline{\includegraphics[width=3.2in]{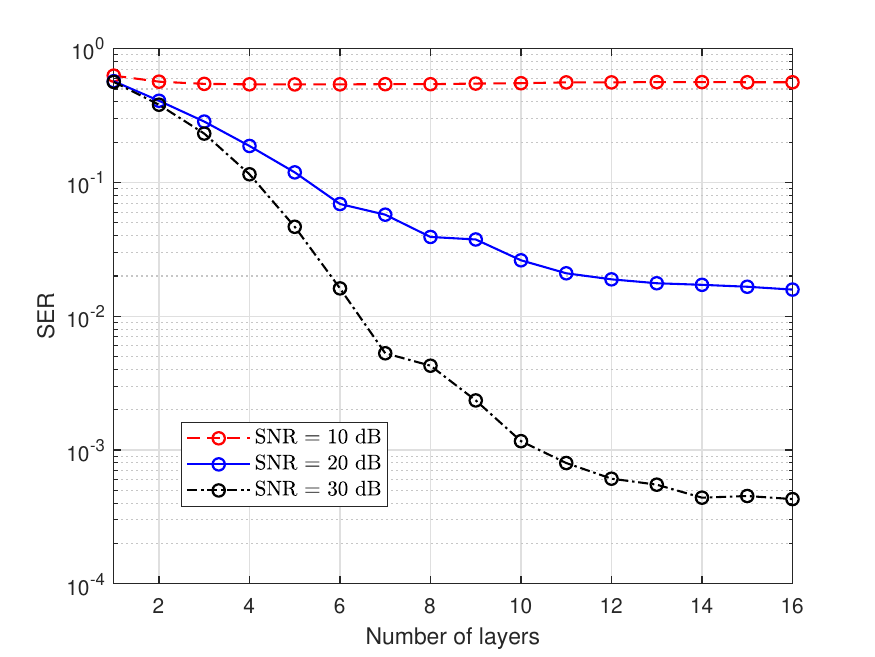}}
  \centerline{(b) QPSK.}
\end{minipage}
\caption{.~~Convergence analysis of the AMP-GNN  versus the
number of layers under QPSK and 16-QAM modulation with $32\times32$ MIMO systems.}
\label{Fig:convergence}
\end{figure}
\subsection{Convergence Analysis}
First, we analyze the convergence of the AMP-GNN network for MIMO detection. Fig.\,\ref{Fig:convergence} illustrates the SER performance versus the number of layers
under various SNRs with QPSK and 16-QAM symbols. The numbers of antennas and users are $M=N=32$. As shown in the figure, the AMP-GNN converges within ten layers  for all the cases. Furthermore, more numbers of layers are required to be convergent in high SNRs and modulation order.
Based on the above observations, we consider the AMP-GNN-based detectors with ten layers ($T = 10$) in following simulation.
\begin{figure}
\begin{minipage}{3in}
  \centerline{\includegraphics[width=3.0in]{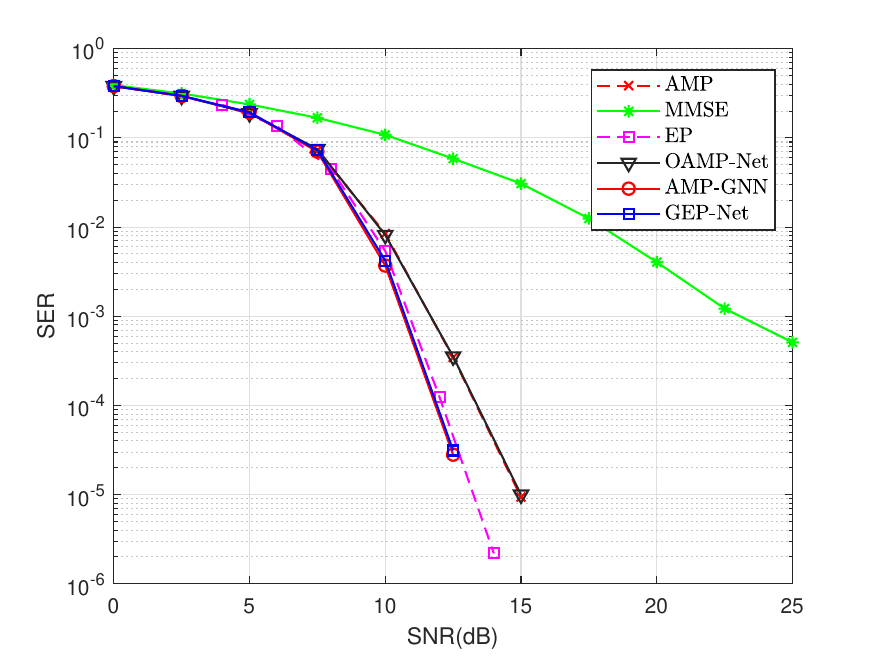}}
  \centerline{(a) QPSK.}
\end{minipage}
\hfill
\begin{minipage}{3in}
  \centerline{\includegraphics[width=3.0in]{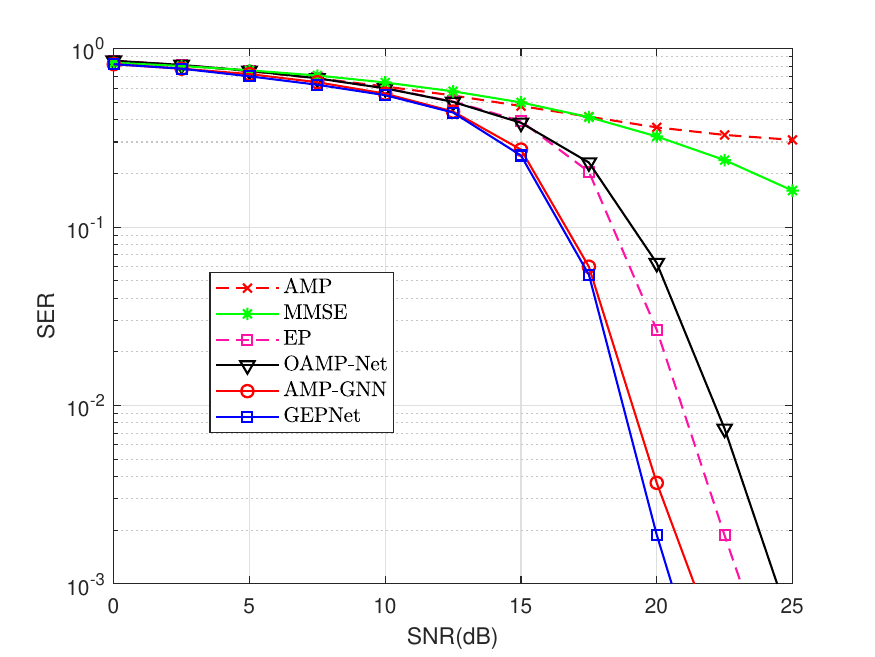}}
  \centerline{(b) 16-QAM.}
\end{minipage}
\caption{.~~SER comparison of the AMP-GNN with other MIMO detectors under $64\times64$ Rayleigh MIMO channels with
QPSK  and 16-QAM symbols.}
\label{Fig:rayleigh}
\end{figure}
\subsection{Performance Comparison}
\begin{figure}
\begin{minipage}{3in}
  \centerline{\includegraphics[width=3.0in]{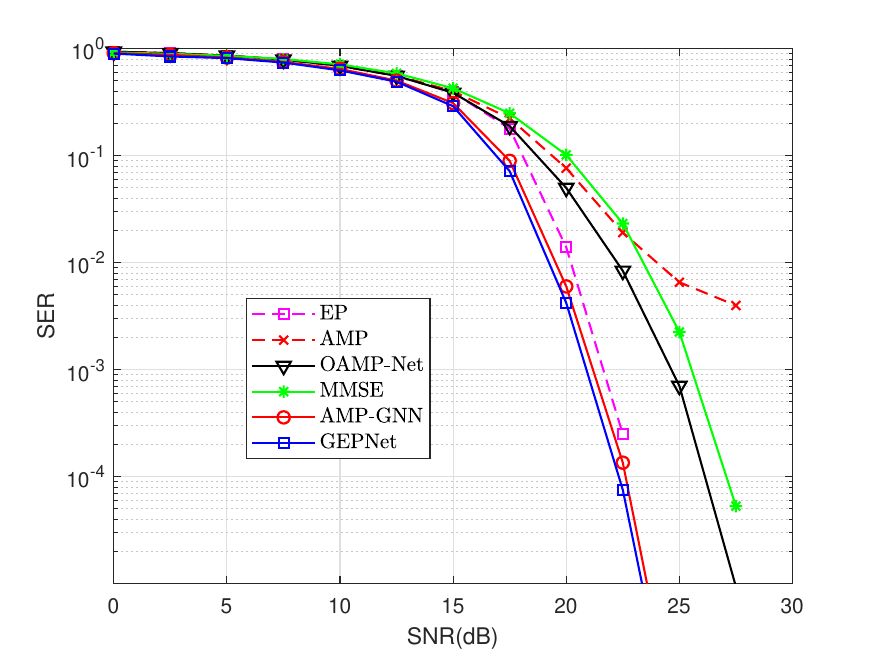}}
  \centerline{(a) $64\times32$ MIMO with $64$-QAM.}
\end{minipage}
\hfill
\begin{minipage}{3in}
  \centerline{\includegraphics[width=3.0in]{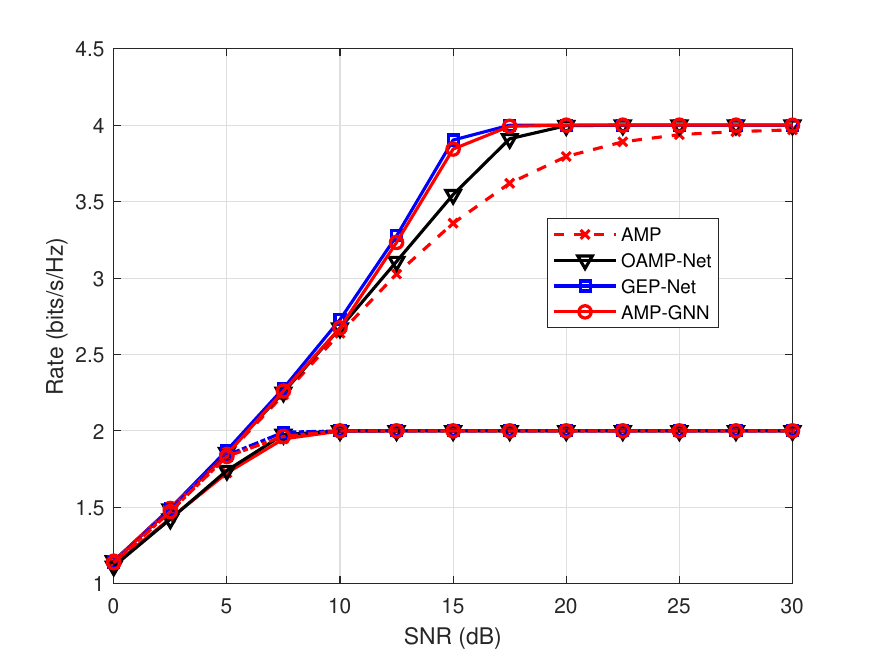}}
  \centerline{(b) $32\times32$ MIMO with QPSK and $16$-QAM.}
\end{minipage}
\caption{.~~SER and achievable rate performance of the AMP-GNN with other MIMO detectors.}
\label{Fig:64QAM}
\end{figure}

Fig.\,\ref{Fig:rayleigh} compares the average SER of the AMP-GNN with those of the baseline detectors. As can be observed from the figure, the AMP-GNN outperforms almost all  MIMO detectors except for the GEPNet detector. In particular, the AMP-GNN outperforms the AMP detector at all SNRs, which demonstrates that the GNN module can improve the AMP detector significantly. Specifically, if we target an SER=$10^{-2}$, then the performance gain is approximately $3.9$ dB compared to the AMP detector. The reason for the performance improvement is that the GNN refine the equivalent AWGN model with a more accurate distribution $\sfp_{\mathrm{GNN}}^{(t)}(x_n=s_{i}|\by)$. Furthermore, the AMP-GNN has only $0.8$ dB performance loss compared to the GEPNet detector in a $4\times4$ MIMO system when we target at an SER=$10^{-2}$. The performance loss then reduces to $0.3$ dB for $32\times32$ MIMO systems as illustrated in Fig.\,\ref{Fig:rayleigh}. A similar conclusion can be obtained with a higher modulation order in Fig.\,\ref{Fig:64QAM}(a). Thus, we conclude that the AMP-GNN has a comparable performance to GEPNet but with a remarkably reduced computational complexity, especially for massive and ultra-massive MIMO systems.
\subsection{Achievable Rates}
 To provide a clearer performance comparison, we present the achievable rates of different detectors for each user with different modulation symbols. This can be achieved because  message passing-based MIMO detectors can decouple the linear mixing model $\mathbf{y} = \mathbf{H}\mathbf{x}+\mathbf{n}$ into a series of independent equivalent AWGN channels given by
\begin{equation}\label{eqAWGN}
    r_{n}^{(t)} = x_{n} + w_{n}^{(t)},
\end{equation}
where $w_{n}^{(t)}\sim \mathcal{N}_{\mathbb{C}}(w_{n}^{(t)};0,\Sigma_n^{(t)})$.  For an arbitrary discrete constellation $\mathcal{S} = \{s_{1}, \ldots, s_{|\mathcal{S}|}\}$ with equal probability $1/|\mathcal{S}|$, we have the performance of $\mathrm{mmse}(\rho)$  given by,

\begin{equation}\label{eqAWGN}
    \mathrm{mmse}(\rho) = 1-\frac{1}{\pi}\int\frac{\sum_{l=1}^{|\mathcal{S}|}s_{l}e^{-|y-\sqrt{\rho}s_l|^{2}}}{|\mathcal{S}| \sum_{l=1}^{|\mathcal{S}|}s_{l}e^{-|y-\sqrt{\rho}s_l|^{2}}}dy.
\end{equation}
where $y = \sqrt{\rho}x+z$. Thanks to the relationship between the mutual information and MMSE in the SISO case \cite{Capacity_AMP}, we have
\begin{equation}\label{eq:MIMO}
\mathcal{R}_{\mathcal{S}}(\rho^{*}) = I(x;\sqrt{\rho^{*}}x+z)= \int_{0}^{\rho^{*}}\mathrm{mmse}(\rho)d\rho
\end{equation}
where $\rho^{*}= 1/\Sigma_n^{(t)}$. It is shown that the capacity of a SISO-AWGN channel equals to the area under $\mathrm{mmse}(\rho)$ from
$\rho = 0$ to $\rho = \rho^{*}$. As illustrated in Fig.\,\ref{Fig:64QAM}(b), the AMP-GNN outperforms OAMP-Net and AMP detectors, and achieves similar performance to GEPNet, which demonstrates that the GNN module can help multi-user interference cancellation.
\subsection{Robustness to Dynamic Numbers of Users}
In Fig.\,\ref{Robustness_2}(a), we train the AMP-GNN in an $32\times16$  and $32\times32$ MIMO systems and test it in a $32\times24$ MIMO system. As shown in the figure, if we target an SER $ =10^{-3}$,  then the AMP-GNN still has a $2.0$ dB performance gain compared with the conventional AMP detector even when tested with different numbers of users. Furthermore, it has a similar performance as the AMP-GNN trained and tested both in the $32\times24$ MIMO system, which indicates that the AMP-GNN has strong robustness to different numbers of users in the deployment stage. This is because the GNN has the permutation equivariance property which makes it robust against dynamic changes in the number of users.    
\subsection{Robustness to Channel Estimator Error}
\begin{figure}
\begin{minipage}{3in}
  \centerline{\includegraphics[width=3.0in]{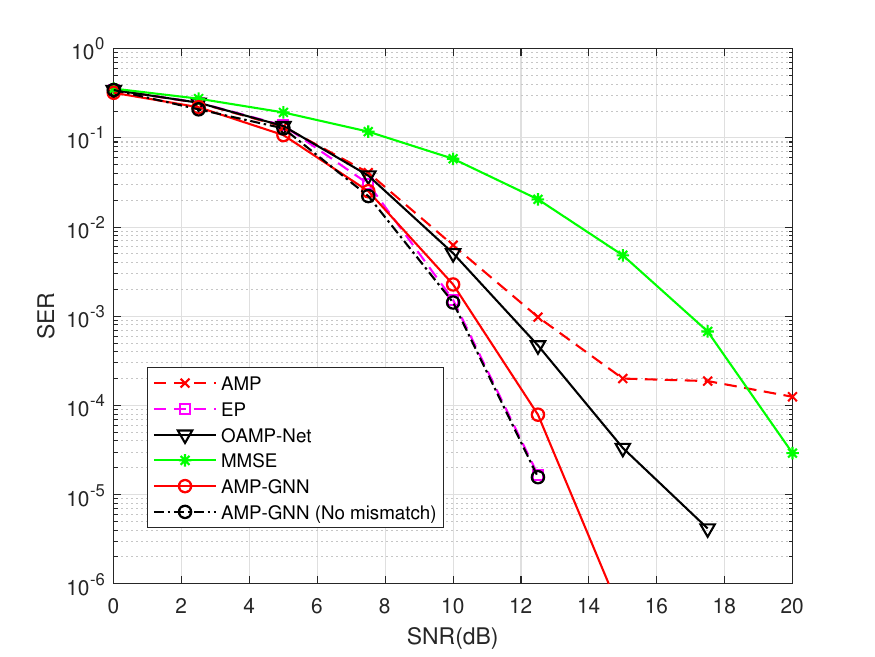}}
  \centerline{(a) Robustness to the number of users.}
\end{minipage}\label{Robustness2a}
\hfill
\begin{minipage}{3in}
  \centerline{\includegraphics[width=3.0in]{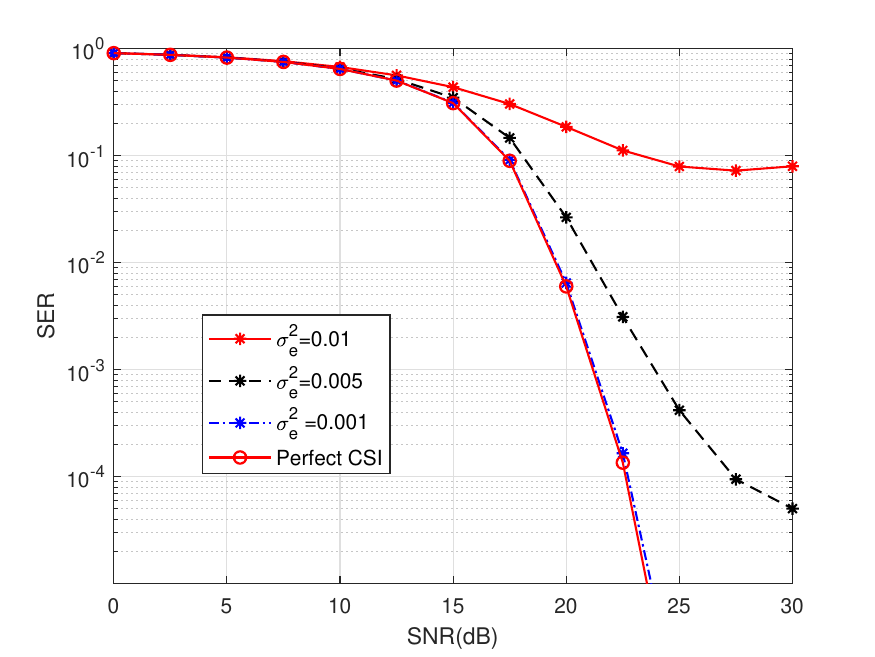}}
  \centerline{(b) Robustness to channel estimator error.}
\end{minipage}\label{Robustness2b}
\caption{.~~Robustness performance of the AMP-GNN to various mismatches.}
\label{Robustness_2}
\end{figure}

In the aforementioned subsections, we assumed AMP-GNN with  perfect CSI. However, channel estimation error normally exists in practical systems even considering high-performance channel estimators. We train the AMP-GNN with perfect CSI and test it with noisy channels, which is given by
\begin{equation}\label{eqHhat}
  \hat{\bH} = \bH + \bE.
\end{equation}
The channel estimator error $\bE \sim \mathcal{N}_{\bbC}(0,\sigma_{e}^{2}\bI)$ and $\sigma_{e}^{2}$ denotes the power of the channel estimator error.
Fig.\,\ref{Robustness_2}(b) shows the performance of the AMP-GNN with various powers of channel estimation error. In particular, the performance of the trained AMP-GNN with $\sigma_{e}^{2}=0.001$ is similar to that with perfect CSI, which demonstrates the AMP-GNN has strong robustness to channel estimator error.

\section{Conclusions}\label{con}
We have developed a novel GNN-enhanced AMP detector for massive/ultra-massive MIMO systems, namely AMP-GNN, which is obtained by incorporating a GNN module into the AMP algorithm. AMP-GNN inherits  the low-complexity of the AMP algorithm and high efficiency of the GNN module. It was shown that  AMP-GNN improves the performance of the AMP algorithm significantly. Simulation results have also demonstrated that the AMP-GNN-based detector has comparable performance as the state-of-the-art GEPNet detector but with a significantly reduced computational complexity. Furthermore, it is robust to the change in the number of users in practical wireless systems. In the future, it will be interesting to apply the AMP-GNN network to other statistical inference problems in wireless communications, such as channel estimation, beamforming design, localization \cite{cooperative_localization}, and nonlinear systems \cite{He_JSTSP}.
\section*{Acknowledgment}
The authors would like to thank Prof. Chao-Kai Wen, from the National
Sun Yat-sen University for the discussion of neural enhanced message passing. Furthermore, the authors would like to thank Prof. Wibowo Hardjawana and Dr. Alva Kosasih from The University of Sydney, for sharing the codes for the GEP-Net.
\appendices
\section{PROOF OF PROPOSITION 1}\label{derivation}
To prove \emph{Proposition 1}, we have the following two Propositions.

\textbf{Proposition 2.} If MPNN and AMP modules in the AMP-GNN are permutation equivariant then the AMP-GNN method is also permutation equivariant.

\emph{Proof:} Due to the transitivity of the permutation equivariance, it suffices to prove  each module is equivariant. Therefore,
 we prove the equivariance for each module separately as follows.

\textbf{Proposition 3.} The AMP algorithm and MPNN are permutation equivariant.

\emph{Proof:}   The AMP algorithm is mainly composed of two modules, the linear and nonlinear modules.  The \emph{linear module} is mainly characterized by Eq.(\ref{Equ:AMP1}-\ref{Equ:AMP4}) and the nonlinear  module is characterized by Eq.(\ref{Equ:AMP5}-\ref{Equ:AMP6}).  As the computation process in Eq.(\ref{Equ:AMP1}-\ref{Equ:AMP4}) is element-wise,
we can obtain $\tilde{\boldsymbol{\Sigma}}^{(t)} = \boldsymbol{\Pi}^{T}\boldsymbol{\Sigma}^{(t)}\boldsymbol{\Pi}$ and $\tilde{\br}^{(t)} = \boldsymbol{\Pi}^{T}\br^{(t)}$ by substituting $\tilde{\bH} = \bH\boldsymbol{\Pi}$. The nonlinear model is also permutation equivariant as
it independently performs the denoising for each user. On the other hand, the MPNN has already been proven to be permutation equivariant \cite{YifeiGNN1}. Thus, all  modules in the proposed AMP-GNN  are permutation equivariant. We can hence conclude that the AMP-GNN is permutation equivariant and robust to the user permutations.

\end{document}